\documentclass{aa}  

\usepackage{graphicx}
\usepackage{txfonts}
\usepackage[T1]{fontenc}
\begin{document}

   \title{Stellar models with self-consistent Rosseland opacities}
   \subtitle{Consequences for stellar structure and evolution}

\titlerunning{Stellar models with self-consistent Rosseland opacities}
   \author{A. Hui-Bon-Hoa
          \inst{1}
          }

   \institute{IRAP, Universit\'e de Toulouse, CNRS, UPS, CNES, Toulouse, France\\
            \email{alain.hui-bon-hoa@irap.omp.eu}
             }

   \date{Received 9 December 2020; accepted 12 January 2021}

 
  \abstract
 {The building of a stellar structure requires knowing the Rosseland mean opacity at each layer of the model. This mean opacity is very often interpolated in pre-computed tables due to the overwhelming time to compute it from monochromatic cross sections. The main drawback to using tables is that the opacities can be inconsistent with the actual local chemical composition, for instance in the regions of the star where nucleosynthesis occurs.}
   {We study the effects of self-consistent Rosseland mean opacity calculations on the stellar structure and evolution, in comparison with models where the metal mixture remains equal to the initial one.} 
   {We   developed a strategy that allows very fast calculations of Rosseland opacities from monochromatic cross sections. We are then able to compute evolutionary tracks with models whose Rosseland opacities are fully consistent with the chemical mix everywhere in the star. This method has been implemented in the Toulouse-Geneva evolution code.}
   {Our self-consistent models show very small structural differences compared to models where the Rosseland opacity is computed with a fixed metal mixture. As a consequence, the main-sequence evolutionary tracks are almost the same for models of mass ranging from 2 to 8~$\mathrm{M_{\odot}}$. At a given surface gravity the relative difference in age is lower than 2~\% and generally below 1~\% between the two kinds of calculations, the self-consistent model being younger most of the time. Unless such a precision in age is sought out, the use of tabulated Rosseland opacities with a metal content defined globally is still acceptable, at least in main-sequence stars where  the chemical mix changes only through nucleosynthesis.}
   {}


   \maketitle
%

\section{Introduction}

The Rosseland mean opacity (RMO) is a major ingredient in the building of stellar structure since it describes how the radiative field carries energy inside the star. In particular, the understanding of pulsating stars depends critically on the accuracy and completeness of the opacity data, so that \cite{Simon1982} entitled his study on Cepheid models `A plea for reexamining heavy element opacities in stars'. \cite{Seaton_etal1994} presented an abstract of the evolution of the opacity data up to that time, and how Simon's plea was answered with the availability of the opacity
data from OPAL \citep{Iglesias_Rogers1996} and the Opacity Project (OP) \citep{Seaton_etal1994,Badnell_etal2005}.\\
From the user's point of view, OPAL provides online\footnote{https://opalopacity.llnl.gov/} pre-computed Rosseland opacity tables for usual mixes, as well as  computation on-the-fly of tables for any user-defined chemical composition. Monochromatic cross sections are available only to certain research groups  \citep[e.g. the Montpellier-Montr\'eal code,][]{Turcotte_etal1998a}. The OP also provides online\footnote{http://opacities.osc.edu/} calculations of Rosseland opacity tables for single mixtures. In addition, the monochromatic data and the codes to handle them \citep{Seaton2005} can also be downloaded. Most of the underlying atomic data are also available through the TOPbase database \citep{Cunto_etal1993}. Sets of opacity data (monochromatic as well as Rosseland means) have also been released by the Los Alamos group\footnote{https://aphysics2.lanl.gov/apps/} \citep[OPLIB;][and references therein]{Colgan_etal2016}.

Many physical processes lead to departures from the chemical composition of the medium where a star forms. For stars in nuclear burning phases, the chemical composition obviously changes in the layers where nucleosynthesis occurs. Atomic diffusion is a transport process that can also give rise to abundance changes: the chemicals move according to the balance between the radiative accelerations and the local gravity. The radiative accelerations being linked to the absorption properties of each kind of ion, each species will have its own abundance profile with respect to the location in the star. Therefore, the chemical composition will no longer be the initial one, nor be homogeneous in layers where the diffusion timescale is short enough compared to the stellar lifetime \citep[e.g.][]{Turcotte_etal1998a,Richer_etal2000,Theado_etal2012}. Accretion (e.g. of circumstellar material, planets) can also alter the chemical composition, at least in the outermost layers \citep{Gonzalez1998}.\\
A self-consistent stellar model requires   computing RMOs according to the local abundance in each part of the star. This computation makes use of the monochromatic cross sections of each element, weighted by its abundance. Several evolution codes can compute RMOs from monochromatic data, at least for some parts of the stellar structure. For instance, STARS \citep{Hu_etal2011}, TGEC \citep{Theado_etal2009}, MESA \citep{Paxton_etal2019}, and CESTAM \citep{Deal_etal2020} use the OP data. Instead, the OPAL data are used in the Montpellier-Montr\'eal code \citep{Turcotte_etal1998a}.

Using  the OP routines to compute Rosseland opacities dynamically requires a huge amount of time, even if they are optimised \cite[e.g. by using parallelisation, as in][]{Hui-Bon-Hoa_Vauclair2018}. Nevertheless, with a suitable computing strategy, we are now able to evolve a star with RMOs that are consistent with the chemical mix everywhere in the star, within a reasonable time.

In this study we  evaluate the changes in the stellar evolution between models computed with self-consistent RMOs and those calculated with fixed metal abundance ratios, mimicking the use of opacity tables specified only by the mass fraction of H, He, and that of the metals. We only address the effect of nucleosynthesis on the composition and assume a homogeneous composition outside the core. In Sect.~\ref{tgec} we describe the evolution code and the approach we adopt for the RMO calculations. In Sect.~\ref{results} we show the evolution of models computed with self-consistent Rosseland opacities compared to models whose metal content ratios are fixed, and discuss the differences induced by the computation method on the stellar parameters.

\section{Numerical models}\label{tgec}

\subsection{Stellar evolution code}
The computations are carried out with the Toulouse-Geneva evolution code \citep[TGEC;][]{Hui-Bon-Hoa2008,Theado_etal2012,Hui-Bon-Hoa_Vauclair2018}. This 1D code implements the nuclear reaction rates from the NACRE compilation \citep{Angulo1999}, and the OPAL2001 equation of state \citep{Rogers_Nayfonov2002}. The energy transport in convective zones is estimated using the mixing-length theory, with a ratio of the mixing length to the pressure scale height equal to 1.8.\\
On-the-fly calculation of the Rosseland opacities using the Opacity Project OPCD v.3.3 data and original routines \citep{Seaton2005} was introduced by \cite{Theado_etal2009}, but only in the outermost regions ($\log T<6.8$), and OPAL tables were used in hotter layers down to the core. In the present study, the RMOs are computed with the OP cross sections throughout the whole star.\\
In this paper we are only interested in the effect of atomic opacities. To avoid the contribution of molecules to the opacity, we    model stars of masses above 2~$\mathrm{M_{\odot}}$ \citep{Alexander_Ferguson1994}.\ Convection is the only transport process. No overshooting is considered around the central convective zone when present.

\subsection{Computational strategy for the Rosseland opacity}

\begin{figure}
   \centering
   \includegraphics[width=.49\textwidth]{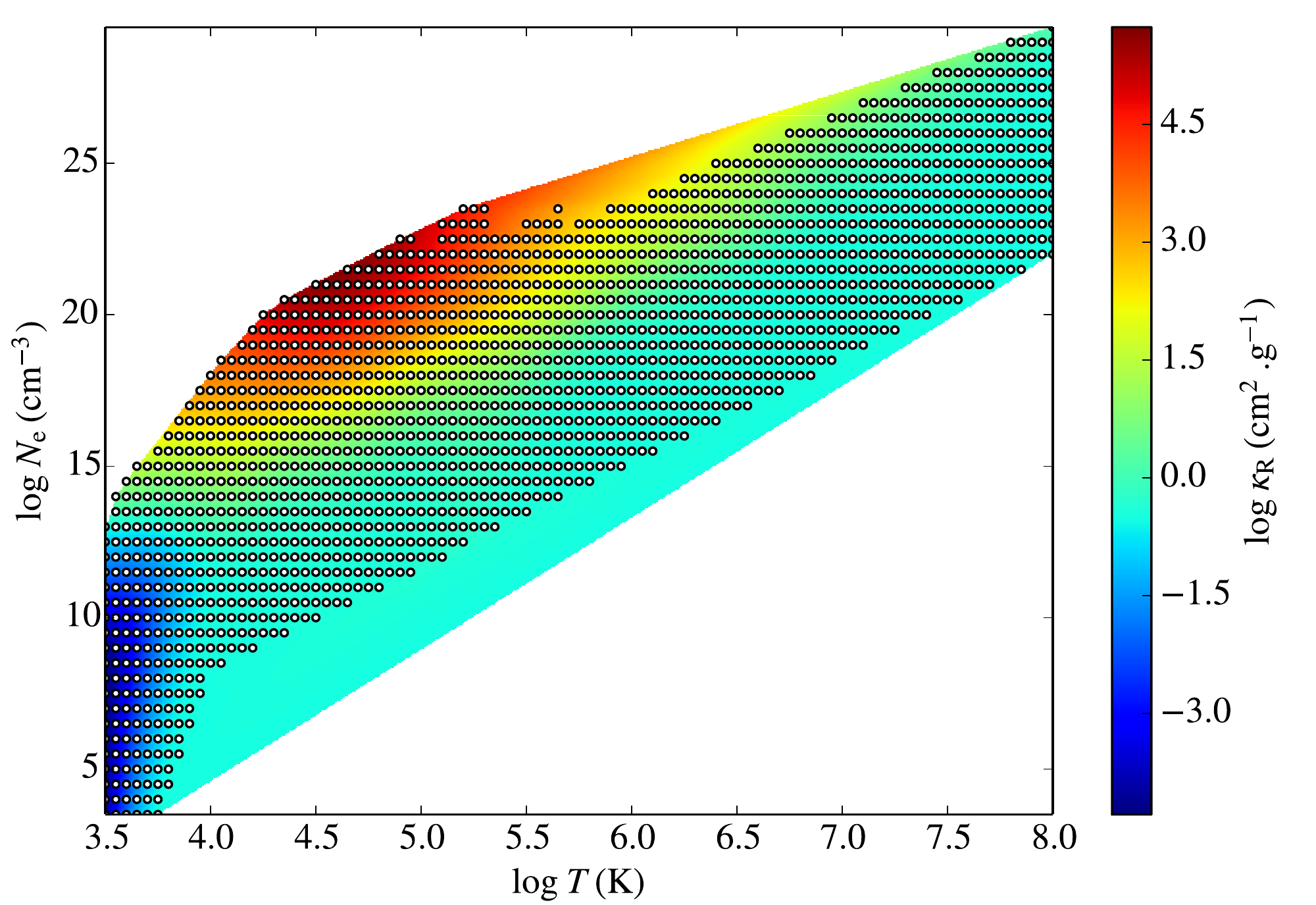}
      \caption{Logarithm of the Rosseland mean opacities $\kappa_\mathrm{R}$ in the OP ($T$, $N_{\mathrm e}$) domain for the adopted solar mix (see text for details). The dots represent the grid points where the OP data are available.}
         \label{figure:OPdomain}
\end{figure}

By definition, the Rosseland mean opacity $\kappa_\mathrm{R}$ is a harmonic mean of the monochromatic opacities $\kappa_\nu$ weighted by the temperature derivative of the Planck function $\frac{dB_\nu}{dT}$,
\begin{equation}
\frac{1}{\kappa_\mathrm{R}}=\frac{\int_{0}^{\infty} \frac{\frac{dB_\nu}{dT}}{\kappa_\nu} \mathrm{d}\nu}{\int_{0}^{\infty} \frac{dB_\nu}{dT}  \mathrm{d}\nu}
,\end{equation}
which can be rewritten as
\begin{equation}\label{eq:RMO}
\frac{1}{\kappa_\mathrm{R}}=\int_{0}^{\infty} \frac{F(u)}{\kappa(u)} \mathrm{d}u
,\end{equation}
where $F(u)$ is the normalised temperature derivative of the Planck function, expressed in terms of $u=\frac{h\nu}{k\mathrm{T}}$ \citep[for its expression, see Eq.~8 of][]{Badnell_etal2005}. $\kappa_\nu$ (alternatively $\kappa(u)$) is equal to the sum of the monochromatic cross sections $\sigma_{A,\nu}$ of each chemical element weighted by its number fraction $N_A$:
\begin{equation}\label{eq:kappa_nu}
\kappa_\nu=\sum_A \sigma_{A,\nu} N_A
.\end{equation}
Equation~\ref{eq:kappa_nu} shows how the RMO depends on the chemical composition. In RMO calculations the computing time is spent in the evaluation of the integral of Eq.~\ref{eq:RMO}.\\
When using the original OPCD routines all the necessary data are read from the disk each time a value of RMO is computed. 
So the first step to reduce computing time is to read only once all the required data at the beginning of an evolution run, and store them in memory.\\
Then, the goal is to perform as few calculations as possible. During the building of a stellar structure, the RMO of each layer depends on the local temperature, electron density, and chemical mix. The RMO is obtained through an interpolation of the surrounding RMOs at the OP grid points in the temperature-electron density ($T$, $N_{\mathrm e}$) plane.  The interpolation uses a bi-cubic scheme (for $T$ on one hand and $N_{\mathrm e}$ on the other) and involves 16 points (4 in each direction). The RMO values at these grid points are computed with the local mix. For the models in this study, most of the layers have a composition equal or very close to the initial one because nucleosynthesis occurs in a small part of the star. Therefore, we compute the RMOs for the initial chemical mix at the beginning of the evolution run for all the ($T$, $N_{\mathrm e}$) domain of OP (a subspace of the rectangle defined by $3.5\leq\log T\leq8$, $3.5\leq\log N_{\mathrm e} \leq29.5$, see Fig.~\ref{figure:OPdomain}), and the resulting table is stored in memory. Wall-clock time is reduced through a parallelisation of this calculation. For each layer involving the initial composition, its RMO is interpolated in this table.\\
If the composition of the current layer differs from the initial one, then the RMOs at the 16 grid points needed for the interpolation are computed from the monochromatic cross sections, with a parallelised routine, and saved in memory. This is useful in the case of neighbouring layers with the same composition but different from the initial one (e.g. in a convective zone); from one layer to the other, the interpolation can require some of the same grid points. If so, the RMO values are then recalled. The threshold of the relative abundance change is set at $10^{-4}$. A test with smaller values showed an increase in  the computing time without any significant impact of the resulting RMO on the stellar evolution.

In the currently available opacity tables the chemical composition is often described only by the mass fractions of H, He, and the metals (as usual, $X$, $Y$, and $Z$); that is, the metals keep their initial ratios (e.g. OPAL type 1 tables). This is the crudest approximation for the metal content when the composition differs from the initial one. To compare our self-consistent opacity calculations with models using such tables (hereafter SC-models and Z-models, respectively), we mimic the last kind of calculations by forcing the metals to keep their initial ratios in our routine. This method has been preferred to the use of tabulated RMOs since only the chemical composition changes in our calculations. We thus avoid any additional numerical effects occurring when creating the tables, and then when they are used, as shown by \cite{D_souza_2014}.

To put figures on our computations, each evolutionary track involves 400 to 500 models, each with 700 to 900 layers. The computing time ranges from 30 to 50~mn using 16 CPUs.


\section{Results and discussion}\label{results}

In this section we present the evolution of stars in the mass range [2;8] $\mathrm{M_{\odot}}$ by steps of 2~$\mathrm{M_{\odot}}$. The models are evolved from the pre-main sequence (PMS) until the hydrogen core exhaustion, defined by a hydrogen mass fraction at the centre lower than $2.10^{-4}$. As we are only interested by the effect of the opacity calculation method during the main sequence (MS), the PMS tracks are the same between the SC- and Z-models. The initial abundances are the solar values according to  \cite{Asplund_etal2009}, with the meteoritic values of \cite{Lodders_Palme_Gail2009} for the refractory elements, as suggested by \cite{Serenelli2010}.

\begin{figure}
   \centering
   \includegraphics[width=.49\textwidth]{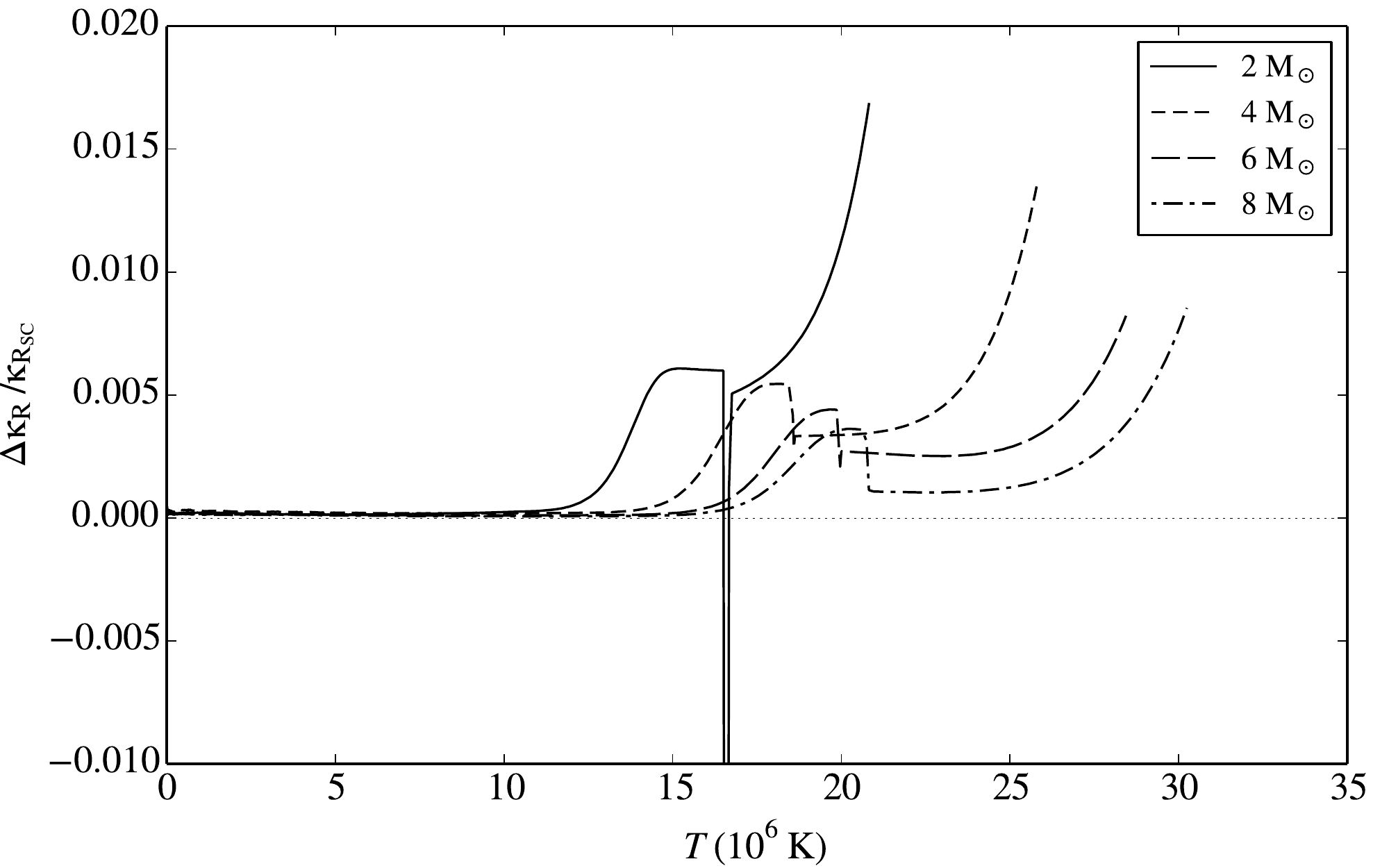}
      \caption{Relative Rosseland mean opacity differences between Z- and SC-models as a function of temperature at the beginning of the main sequence. The discontinuities between $T=1.6~10^7$ and $2.1~10^7$ K are due to the shift of the central convective zone boundary between the two kinds of models.}
         \label{figure:diffOpacs}
\end{figure}

\begin{figure}
   \centering
   \includegraphics[width=.49\textwidth]{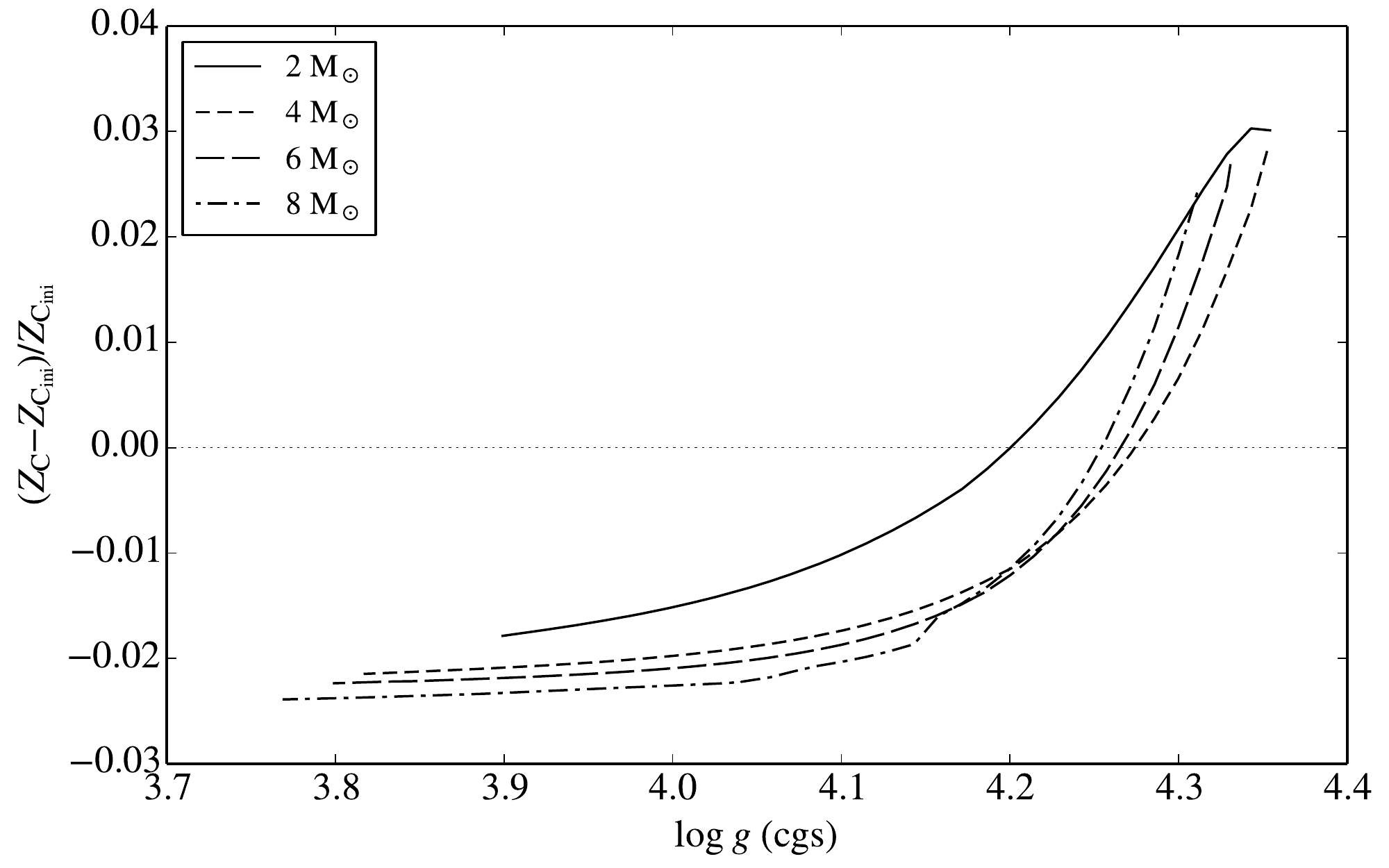}
      \caption{Relative difference between the central metal mass fraction $Z_{\rm C}$ and its initial value $Z_{\rm C_{ini}}$ {vs.} $\log g$. At the beginning of the MS (i.e. at high $\log g$ values) the difference with respect to the initial value $Z_\mathrm {C_{ini}}$ is due to the nucleosynthesis occurring during the pre-main sequence.}
         \label{figure:Z}
\end{figure}

\begin{figure}
   \centering
   \includegraphics[width=.49\textwidth]{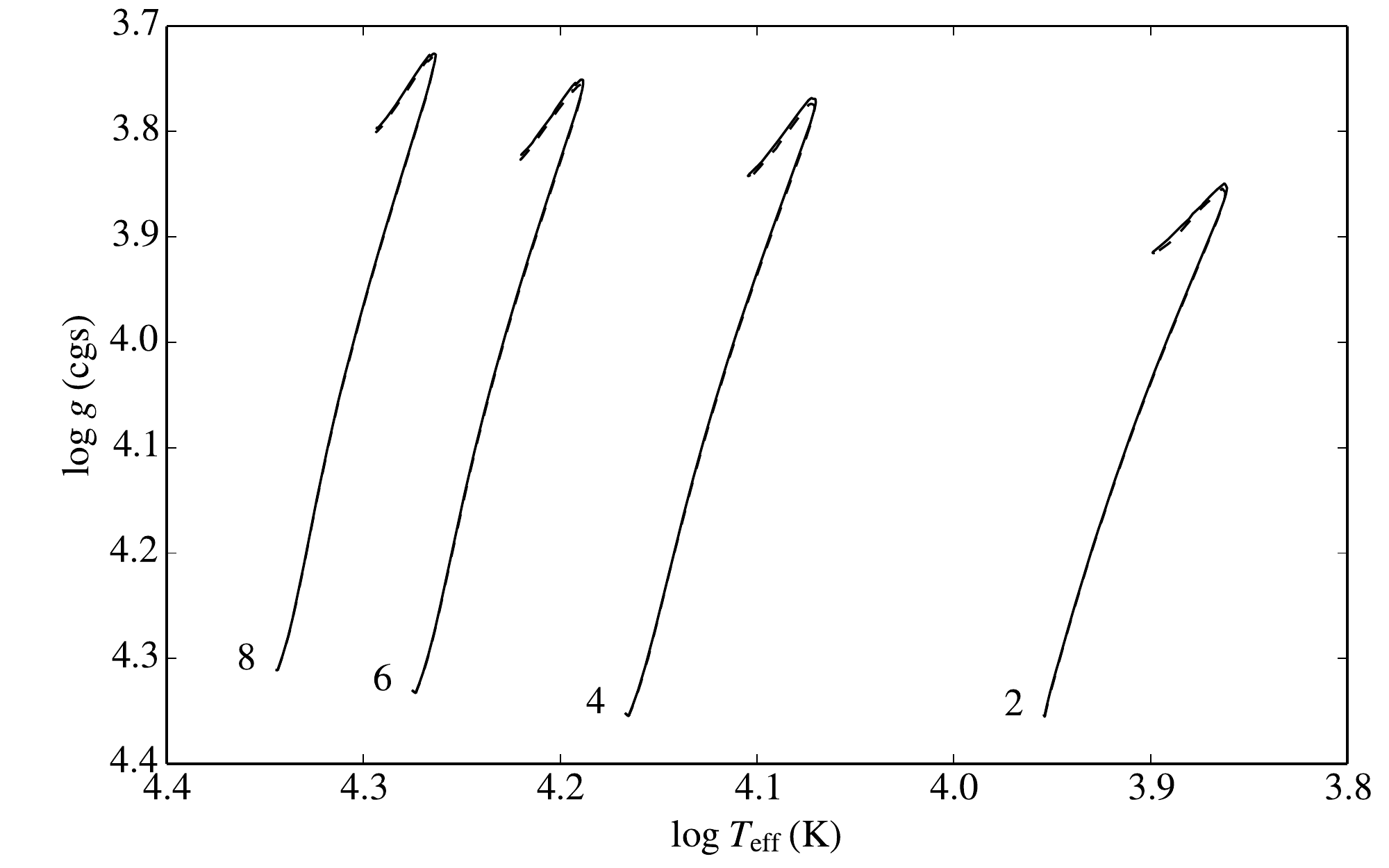}
      \caption{Evolutionary tracks for SC-models (solid lines) and Z-models (dashed lines) in the ($\log T_{\mathrm{eff}}, \log g$) plane. The stellar masses (in $\mathrm{M_{\odot}}$) are shown at the beginning of each pair of tracks.}
         \label{figure:HR}
\end{figure}

The relative difference in Rosseland opacity between the Z-models ($\mathrm{\kappa_{R_Z}}$) and SC-models ($\mathrm{\kappa_{R_{SC}}}$) is defined by
\begin{equation}\notag
\mathrm{\frac{\Delta \kappa_R}{\kappa_{R_{SC}}}=\frac{\kappa_{R_Z}-\kappa_{R_{SC}}}{\kappa_{R_{SC}}}},
\end{equation}
and is plotted against temperature in Fig.~\ref{figure:diffOpacs} for the four models at the beginning of the MS. The difference in opacity between the Z- and SC-models is   most important in the core where nucleosynthesis has already changed the chemical composition during the PMS. At the beginning of the MS, a greater RMO for the Z-models lies in the increase in $Z$ compared to the initial metal content combined to the metal composition; in the SC-models the $Z$ variation reflects the changes in mass fraction of the various isotopes of C, N, and O, whereas the increase in $Z$ is shared by all the metals in the Z-models. In particular, the mass fraction of Fe is greater in the Z-models. Iron being  the major contributor to the RMO among the metals around the stellar centre, the mean opacity increase is mainly due to its higher abundance in the Z-models. In more detail, the shape of the relative opacity difference {versus} $T$ is caused by a greater increase in the contribution of Fe at the centre than at the convective edge in the Z-models, related to the dependence on temperature and electronic density of the monochromatic cross sections.\\
Figure~\ref{figure:Z} shows the evolution of $Z$ as a function of $\log g$. As a consequence of the PMS nucleosynthesis, $Z$ has already departed from its initial value at the beginning of the MS, and is larger with decreasing stellar mass in our mass range, explaining the greater mean opacity in the stellar core as the mass of the star is smaller (see Fig.~\ref{figure:diffOpacs}).
As the star evolves, $Z$ decreases in the central parts due to the abundance variations of the CNO isotopes as the nucleosynthesis proceeds, and the opacity difference decreases as well. It even changes its sign so that the reverse effect appears; the central opacity is greater in the SC-models than in the Z-models.

\begin{figure}
   \centering
   \includegraphics[width=.49\textwidth]{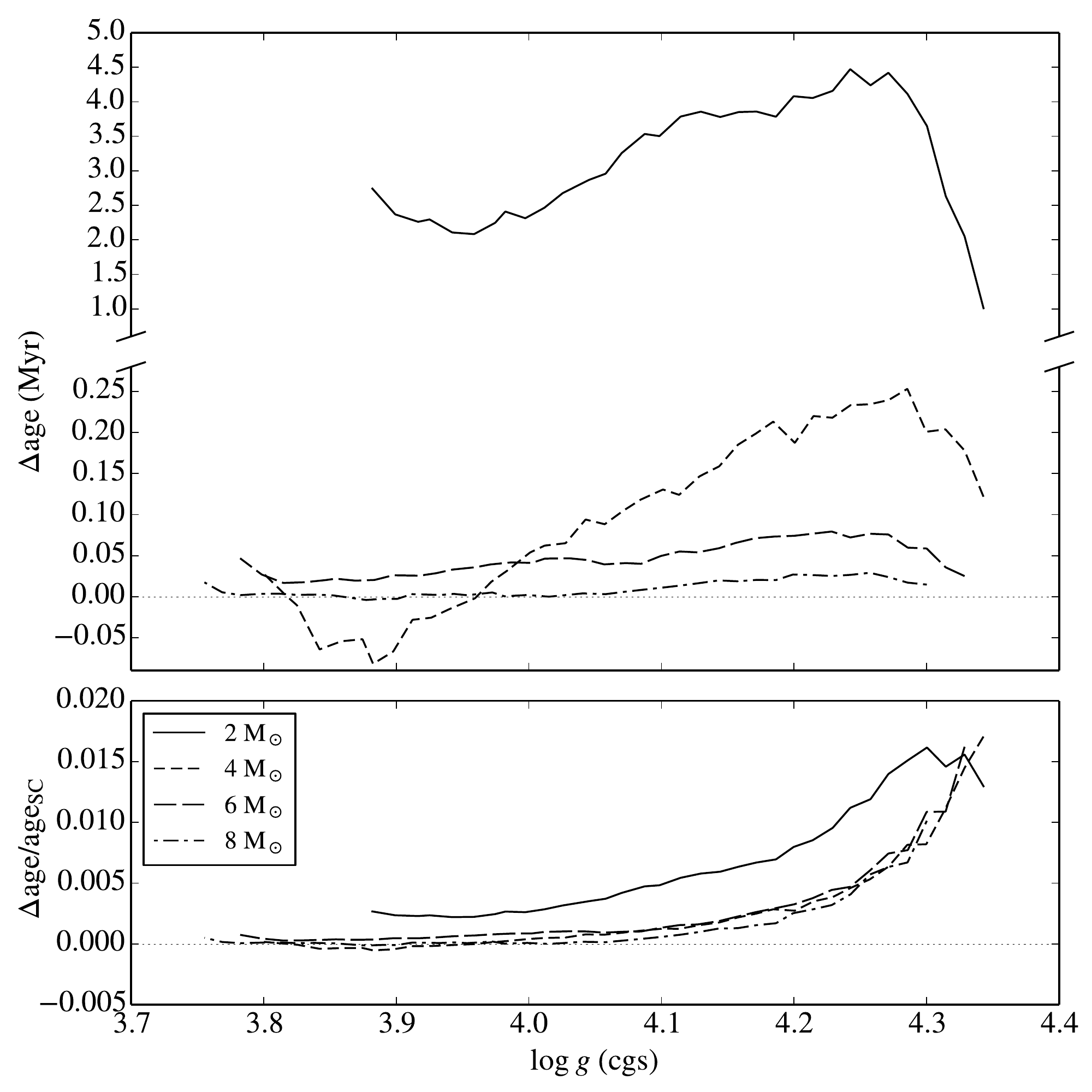}
      \caption{Age difference (upper panel) and relative age difference (lower panel) between the  Z- and SC-models {vs.} $\log g$. The age difference is non-zero between the first Z- and SC-model because their structures are slightly different with a larger $\log g$ for the Z-model.}
         \label{figure:ageDifference}
\end{figure}

Figure~\ref{figure:HR} presents the MS evolutionary tracks for both types of models in the ($\log T_{\mathrm{eff}}, \log g$) plane. Except at the end of the MS, they are almost indistinguishable at given stellar mass. We can then wonder about the age of our models at the same location in this diagram. The upper panel of Fig.~\ref{figure:ageDifference} shows the age difference between the Z- and SC-models as a function of $\log\ g$. At the beginning of the MS the Z-models are older than the SC-models at a given $\log\ g$.  This difference decreases as the stars evolve, and can even have the opposite sign for some models. This age difference is due to the core opacity. In our mass range part of the core is convective, and a higher opacity leads to a bigger convective core. This means more nuclear fuel for a similar stellar structure, and the evolution is therefore slower. At the beginning of the MS the Z-models have a higher core opacity, and the increase in the convective core mass amounts from 1~\% up to 3~\% for the 8 and 2~$\mathrm M_\odot$ models, respectively. The age difference with respect to the SC-models then increases, with a larger difference for lower masses because of a bigger convective core. Later on the core opacity difference reduces and changes its sign. The subsequent variation in the age difference is negative so that the age difference decreases, and can even be of the opposite sign (Z-model younger than SC-model) for the 4 and 8~$\mathrm M_\odot$ models at the end of the MS.

The relative age difference is expressed as
\begin{equation}\notag
\mathrm{\frac{\Delta age}{age_{SC}}=\frac{age_Z-age_{SC}}{age_{SC}}},
\end{equation}
where $\mathrm{age_{SC}}$ and $\mathrm{age_{Z}}$ are respectively the age of the SC-model and of the Z-model.
Its behaviour {versus} $\log g$ is shown in the lower panel of  Fig.~\ref{figure:ageDifference}. The relative difference in age (in absolute value) does not exceed 2~\%, and is, for  most of the MS lifetime, below 1~\% for the 4 to 8~$\mathrm M_\odot$ models. This value is much smaller compared to those due to changes in the input physics of the model, in particular atomic diffusion with radiative accelerations, which amounts to 9~\% for a 1.4~$\mathrm M_\odot$ star \citep{Deal_etal2018}.

As a test of the influence of the initial composition, we evolved the 2~$\mathrm{M_{\odot}}$ model with the \cite{Grevesse_Noels1993} mix and came up with the same conclusions as for the \cite{Asplund_etal2009} composition.

\section{Conclusion}\label{discussion}

Having the ability to compute the Rosseland opacities from monochromatic data in a reasonable time, we show that the evolutions of model stars between 2 and 8 $\mathrm {M_\odot}$ are very similar whether we take into account the details of the metal content in the stellar core or not, the evolutionary tracks being superimposed during almost all the MS. Considering the detailed core composition would affect the age determination to less than 2\% in the worst case. Therefore, calculations with tabulated RMOs in terms of $X$, $Y$, and $Z$ still provide reliable models if this level of  precision is not required. Further work is underway to address other mass ranges, in particular for lower masses.

If the metal content departs from the initial composition in the envelope of an MS star, the impact of a self-consistent computation of the Rosseland opacity should remain small as long as the metals have a much smaller contribution to the opacity compared to H or He in the layers of concern. This is not the case around $\log T=5.3$, where the iron-group elements dominate the Rosseland opacity in the so-called $Z$-bump. Departures from the initial metal composition in this region can be due to atomic diffusion whenever hydrodynamical processes are weak enough for abundance stratifications to exist. This can also happen with the accretion of matter with a metal mixture different from that of the star if the accreted material reaches these layers. Self-consistent RMO calculations are then mandatory in order to have a stellar structure in agreement with the local chemical composition. This justifies some hybrid implementations of RMO calculations using on-the-fly computations in the upper envelope and opacity tables in the central parts, as in CESTAM \citep{Deal_etal2020} and in previous versions of TGEC \citep{Theado_etal2009}.

\begin{acknowledgements}
I am grateful to the anonymous referee whose comments and remarks helped to improve the manuscript. This work was supported by the "Programme National de Physique Stellaire" (PNPS) of CNRS/INSU co-funded by CEA and CNES.
\end{acknowledgements}

%
%

   \bibliographystyle{aa} 
   \bibliography{biblio}

\end{document}